\documentclass[12pt]{article}
\usepackage{a4wide}
\usepackage{amsmath}
\usepackage{graphicx}
\usepackage{bm}

\newcommand{\vv}{{\bm v}}

\newcommand{\ww}{{\bf w}}

\newcommand{\bu}{\ensuremath{\mathbf{u}}}
\newcommand{\bx}{\ensuremath{\mathbf{x}}}

\newcommand{\bnabla}{\ensuremath{\boldsymbol{\nabla}}}

\newcommand{\ff}{{\mbox{$\mathbf{ f}$}}}

\newcommand{\bDelta}{{\boldsymbol{\Delta}}}

\providecommand\bnabla{\boldsymbol{\nabla}}

\newcommand{\gggg}{\mbox{\bf g}}
\newcommand{\ee}{\mbox{\bf e}}
\newcommand{\cc}{\mbox{\bf c}}
\newcommand{\mm}{\mbox{\bf m}}
\newcommand{\vel}{\mbox{\boldmath$c$}}
\newcommand{\uu}{\mbox{\boldmath$u$}}
\newcommand{\xx}{\mbox{\boldmath$x$}}

\newcommand{\KK}{\mbox{\boldmath$K$}}
\newcommand{\LL}{\mbox{\boldmath$L$}}

\newcommand{\CC}{\mbox{\boldmath$C$}}

\newcommand{\bONE}{\mbox{\boldmath$1$}}

\begin{document}


\title{Entropic lattice Boltzmann method for microflows}
\author{S. Ansumali, I. V. Karlin\thanks{Corresponding author}, C. E. Frouzakis, K. B.
Boulouchos
\bigskip\\
  Aerothermochemistry and Combustion Systems Laboratory\\
  Swiss Federal Institute of Technology \\ CH-8092 Zurich, Switzerland}


\date{\today}

\maketitle

\begin{abstract}

A new method for the computation of flows at the micrometer scale
is presented. It is based on the recently introduced minimal
entropic kinetic models. Both the thermal and isothermal families
of minimal models are presented, and the simplest isothermal
entropic lattice Bhatnagar-Gross-Krook (ELBGK) is studied in
detail in order to quantify its relevance for microflow
simulations. ELBGK is equipped with boundary conditions which are
derived from molecular models (diffusive wall). A map of
three-dimensional kinetic equations onto two-dimensional models is
established which enables two-dimensional simulations of
quasi-two-dimensional flows. The ELBGK model is studied
extensively in the simulation of the two-dimensional Poiseuille
channel flow. Results are compared to known analytical and
numerical studies of this flow in the setting of the
Bhatnagar-Gross-Krook model. The ELBGK is in quantitative
agreement with analytical results in the domain of weak
rarefaction (characterized by Knudsen number ${\rm Kn}$, the ratio
of mean free path to the hydrodynamic scale), up to ${\rm Kn}\sim
0.01$, which is the domain of many practical microflows. Moreover,
the results qualitatively agree throughout the entire Knudsen
number range, demonstrating Knudsen's minimum for the mass flow
rate at moderate values of ${\rm Kn}$, as well as the logarithmic
scaling at large ${\rm Kn}$. The present results indicate that
ELBM can complement or even replace computationally expensive
microscopic simulation techniques such as kinetic Monte Carlo
and/or molecular dynamics for low Mach and low Knudsen number
hydrodynamics pertinent to microflows.
\end{abstract}


\section{Introduction}

Gas flows at the micrometer scale constitute a major portion of
contemporary fluid dynamics of engineering interest. Because of
its relevance to the engineering of micro electro-mechanical
systems (MEMS), the branch of computational fluid dynamics focused
on micro scale phenomena is often called ``microfluidics"
\cite{REVMEMS,Karniadakis2}.

Microflows are characterized by the Knudsen number, ${\rm Kn}$,
which is defined as the ratio of the mean free path of molecules
$\lambda$ and the characteristic scale $L$ of variation of
hydrodynamic fields (density, momentum, and energy). For typical
flows in microdevices, ${\rm Kn}\sim \lambda/L$ varies from ${\rm
Kn}\ll 1$ (almost-continuum flows) to ${\rm Kn} \sim 1$ (weakly
rarefied flows).  Another characteristic property of microflows is
that they are highly subsonic, that is, the characteristic flow
speed is much smaller than the speed of sound. This feature is
characterized by the Mach number, ${\rm Ma}\sim u/c_{\rm s}$,
where $u$ is the characteristic flow speed, and $c_{\rm s}$ is the
(isentropic) speed of sound. Thus, for microflows, ${\rm Ma}\ll
1$.  To be more specific, typical flow velocities are about $0.2$
m/s, corresponding to ${\rm Ma} \sim 10^{-4}$, while values of the
Knudsen number range between $10^{-4}\le Kn \le 10^{-1}$. Finally,
in the majority of applications, microflows are
quasi-two-dimensional.

Theoretical studies of gas flows at finite Knudsen number have
begun several decades ago in the realm of the Boltzmann kinetic
equation. To that end, we mention pioneering contributions by
Cercignani, Sone, and others \cite{Cerci,Sone}. These studies
focused on obtaining either exact solutions of the stationary
Boltzmann kinetic equation, or other model kinetic equations in
relatively simple geometries (most often, infinite or
semi-infinite rectangular ducts), or asymptotic expansions of
these solutions.

While analytical solutions are important for a qualitative
understanding of microflows, and also for the validation of
numerical schemes, they certainly do not cover all the needs of
computational fluid dynamics of practical interest. At present,
two CFD strategies for microflows are well established.

\begin{itemize}
\item  {\it Equations of continuous fluid mechanics with slip
boundary conditions.} The simplest semi-phenomenological
observation about microflows is the break down of the  no-slip
boundary condition of fluid mechanics with increasing Knudsen
number. Since  microflows are highly subsonic, this leads to the
simplest family of models, equations of incompressible or
compressible fluid dynamics supplemented by slip velocity boundary
conditions (a review can be found in \cite{Karniadakis2}). This
approach, although widely used at the early days of microfluidics,
remains phenomenological. Moreover, it fails to predict phenomena
such as non-trivial pressure and temperature profiles observed by
more microscopic approaches.

\item {\it Direct simulation of the Boltzmann kinetic equation.}
On the other extreme, it is possible to resort to a fully
microscopic picture of collisions, and to use a molecular dynamics
approach or a simplified version thereof - the Direct Simulation
Monte Carlo method of Bird (DSMC) \cite{Bird}. DSMC is sometimes
heralded as the method of choice for simulation of the Boltzmann
equation, and it has indeed proven to be robust in supersonic,
highly compressible flows with strong shock waves. However, the
highly subsonic flows at small to moderate Knudsen number is not a
``natural" domain for the DSMC simulations where it becomes
computationally intensive \cite{DSMC}.

\end{itemize}

Since semi-phenomenological computations are not reliable, and the
fully microscopic treatment is not feasible, the approach to CFD
of microflows  must rely on reduced models of the Boltzmann
equation. Two classical routes of reducing the kinetic equations
are well known, the Chapman-Enskog method and Grad's moment method
(for a modern summary and extensions of these methods see, for
example, \cite{GKbook}). The Chapman-Enskog method extends the
hydrodynamic description (compressible Navier-Stokes equations) to
finite ${\rm Kn}$ in the form of a Taylor series, leading to
hydrodynamic equations of increasingly higher order in the spatial
derivatives (Burnett's hydrodynamics). Grad's method extends the
hydrodynamic equations to a closed set of equations including
higher-order moments (fluxes) as independent variables. Both
methods are well suited for theoretical studies of microflows. In
particular, as was already noted by Grad \cite{GradH}, moment
equations are especially well suited for low Mach number flows.

However, applications of Grad's moment equations or of Burnett's
hydrodynamics (or of existing extensions and generalizations
thereof) to CFD of microflows are limited at present because of
several reasons. The most severe difficulty is in formulating the
boundary conditions at the reduced level. Although some studies of
boundary conditions for moment systems were initiated recently
\cite{GrmelaKarZmi}, this problem is far from solved. The crucial
importance of the boundary condition for microflows is actually
expected. Indeed, as the rarefaction is increasing with ${\rm
Kn}$, the contribution of the bulk collisions becomes less
significant as compared to the collisions with the boundaries, and
thus the realistic modelling of the boundary conditions becomes
increasingly important.

In this paper we set up a novel approach to the CFD of microflows.
It is based on the recently developed Entropic Lattice Boltzmann
Method
(ELBM)\cite{ELB1,ELB2,AK1,AK2,AK3,AK4,AK5,Boghosian,RMP,AS}. The
choice of the ELBM for microflows is motivated by two reasons:

\begin{itemize}

\item ELBM is an unconditionally stable simulation method for
flows at low Mach numbers.

\item In contrast to Grad's method, ELBM is much more compliant
with the boundary conditions. Recently, an appropriate boundary
condition for the ELBM was found upon a discretization of the
diffusive wall boundary condition of the Boltzmann equation
\cite{AK4}. This boundary condition was also rediscovered in
\cite{ELBMMICRO}, where ELBM simulations were tested against
molecular dynamic simulations with a good agreement.

\end{itemize}

It should be mentioned that the predecessor of ELBM, the lattice
Boltzmann method (LBM) \cite{Succi}, was employed several times
for microflow simulations \cite{NIE,LIM,SLIP, KWOK}. The
motivation of most of these works was the velocity slip observed
in the LBM simulations using the so-called bounce-back boundary
condition \cite{NIE,LIM,SLIP,KWOK}. Hovewer, since the bounce-back
boundary condition is completely artificial, the results are
questionable \cite{ELBMMICRO}.

The outline of the present paper is as follows: In section
\ref{ELBMsec}, for the sake of completeness,  we present the
general description of isothermal and thermal entropic lattice
Boltzmann models. Then, we describe the ELBM setup for the
simplest situation of isothermal flows, the entropic lattice
Bhatnagar-Gross-Krook model (ELBGK). The case of isothermal models
is important in itself since it allows to study nontrivial flow
phenomena such as velocity slip at the wall, and the well-known
Knudsen minimum of the mass flowrate in pressure-driven channel
flows. Prediction of the Knudsen minimum is a classical benchmark
problem for the simulation of microflows. In section
\ref{BCsection}, we describe in detail the implementation of the
diffusive boundary conditions for the ELBM for two-dimensional
simulations. A separate section \ref{mapsection} is devoted to the
question of how to simulate quasi-two-dimensional flows with
two-dimensional models, and how to map the parameters of the model
onto experimental data and more microscopic simulations. In
section \ref{Simsec}, we present simulation results for the
quasi-two-dimensional Poiseuille flow, and discuss Knudsen's
minimum, comparing results with known asymptotic and analytic
solutions to the Boltzmann kinetic equation. Results are discussed
in section \ref{conclusion}, where we define the domain of
validity of ELBM for microflows. We also suggest a straightforward
application of ELBM results to accelerate more microscopic
simulation approaches like the DSMC method.

\section{Minimal kinetic models} \label{ELBMsec}

We start with a generic discrete velocity kinetic model. Let
$f_i(\xx,t)$ be populations of the $D$-dimensional discrete
velocities $\vel_i$, $i=1,\dots,n_{\rm d}$, at position $\xx$ and
time $t$. The hydrodynamic fields are the linear functions of the
populations, namely
\begin{equation}\label{fielshyd}
\sum_{i=1}^{n_{\rm d}} \{ 1,\, \vel_{i},\, c_i^2 \} f_i =\{\rho,\,
\rho \uu, \, \rho DT+ \rho u^2 \},
\end{equation}
where $\rho$ is the mass density of the fluid, $\rho \uu$ is the
$D$-dimensional momentum density vector, and $e=\rho D T+ \rho
u^2$ is the energy density.
In the case of isothermal
simulations, the set of independent hydrodynamic fields contains
only the mass and momentum densities. It is convenient to
introduce $n_{\rm d}$-dimensional population vectors $\ff$, and
the standard scalar product, $\langle
\ff|\gggg\rangle=\sum_{i=1}^{{n_{\rm d}}}x_iy_i$. For example, for
almost-incompressible hydrodynamics (leaving out the energy
conservation), the locally conserved density and momentum density
fields are written as
\begin{equation}\label{Hyd}
\langle \bONE|\ff\rangle=\rho,\ \langle
\cc_{\alpha}|\ff\rangle=\rho u_{\alpha}.
\end{equation}
Here $\bONE=\{ 1\}_{i=1}^{n_{\rm d}}$,
$\cc_{\alpha}=\{c_{i\alpha}\}_{i=1}^{n_{\rm d}}$, and
$\alpha=1,\dots,D$, where $D$ is the spatial dimension.

The construction of the kinetic simulation scheme begins with
finding a convex function of populations $H$ (entropy function),
which satisfies the following condition: If $\ff^{\rm
eq}(\rho,\uu)$ (local equilibrium) minimizes $H$ subject to the
hydrodynamic constraints (equations (\ref{fielshyd}) or
(\ref{Hyd})), then $\ff^{\rm eq}$ also satisfies certain
restrictions on the higher-order moments. For example, the
equilibrium stress tensor must respect the Galilean invariance,
\begin{equation}
\label{GI}
  \sum_{i = 1}^{{n_{\rm d}}} c_{i\alpha}c_{i\beta}f^{\rm eq}_i(\rho,\uu)=
\rho c_{\rm s}^2\delta_{\alpha\beta}+\rho u_{\alpha}u_{\beta}.
\end{equation}
The corresponding entropy functions for the isothermal and the
thermal models were found in \cite{DHT,AK4,AK5}, and are given
below (see section \ref{Hfunc} and Table \ref{Tab: DiscV}). For
the time being, assume that the convex function $H$ is given.

The next step is to obtain the set of kinetic equations,
\begin{equation}
\label{LBMcont} \partial_t f_i+c_{i\alpha}\partial_{\alpha}f_i=
\Delta_i.
\end{equation}
Let $\mm_1,\dots,\mm_{n_{\rm c}}$ be the ${n_{\rm d}}$-dimensional
vectors of locally conserved fields, $M_i=\langle
\mm_i|\ff\rangle$, $i=1,\dots,{n_{\rm c}}$,
 $n_{\rm c}<{n_{\rm d}}$. The ${n_{\rm d}}$-dimensional vector function $\bDelta$
(collision integral), must satisfy the conditions:
\[ \langle \mm_i|\bDelta\rangle=0,\ i=1,\dots,n_{\rm c}\
(\rm{local\ conservation\ laws}),\]
\[ \langle \bnabla H|\bDelta\rangle\le 0\
(\rm{entropy\ production\ inequality}),\]
where $\bnabla H$ is the row-vector of partial derivatives
$\partial H/\partial f_i$.
Moreover, the local equilibrium vector $\ff^{\rm eq}$ must be the
only zero point of $\bDelta$, that is, $\bDelta(\ff^{\rm
eq})=\mathbf{0}$, and, finally, $\ff^{\rm eq}$ must be the only
zero point of the local entropy production, $\sigma(\ff^{\rm
eq})=0$. Collision integrals which satisfies all these requirements
are called admissible. Let us discuss several possibilities of
constructing admissible collision integrals.

\subsubsection{BGK model}

Suppose that the entropy function $H$ is known. If, in addition,
the local equilibrium is also known as an explicit function of the
locally conserved variables (or some reliable approximation of this
function is known), the simplest option is to use the
Bhatnagar-Gross-Krook (BGK) model. In the case of isothermal
hydrodynamics, for example, we write
\begin{equation}\label{BGKcoll}
\bDelta=-\frac{1}{\tau}(\ff-\ff^{\rm eq}(\rho(\ff), \uu(\ff))).
\end{equation}
The BGK collision operator is sufficient for many applications.
However, it becomes advantageous only if the local equilibrium is
known in a closed form. Unfortunately, often only the entropy
function is known but not its minimizer. For these cases one
should construct collision integrals  based solely on the
knowledge of the entropy function. We present here two particular
realizations of the collision integral based on the knowledge of
the entropy function only.

\subsubsection{Quasi-chemical model}

For a generic case of $n_{\rm c}$ locally conserved fields, let
$\gggg_s$, $s=1,\dots, {n_{\rm d}}-{n_{\rm c}}$, be a basis of the
subspace orthogonal (in the standard scalar product) to the
vectors of the conservation laws. For each vector $\gggg_{s}$, we
define a decomposition $\gggg_{s}=\gggg_{s}^+-\gggg_{s}^-$, where
all components of vectors $\gggg_{s}^{\pm}$ are nonnegative, and
if $g_{si}^{\pm}\ne0$, then $g_{si}^{\mp}=0$. Let us consider the
collision integral of the form:
\begin{equation}
\label{lbMDD} \bDelta=\sum_{s=1}^{{n_{\rm d}}-{n_{\rm
c}}}\varphi_{s}\gggg_{s} \left\{ \exp\left(\langle\bnabla
H|\gggg_{s}^-\rangle\right)- \exp\left(\langle\bnabla
H|\gggg_{s}^+\rangle\right) \right\}.
\end{equation}
Here $\varphi_{s}>0$. By construction, the collision integral
(\ref{lbMDD}) is admissible. If the entropy function is
Boltzmann--like, and the components of the vectors $\gggg_{s}$ are
integers, the collision integral assumes the familiar
Boltzmann--like form. An example of such a collision term for the
$D2Q9$-discrete velocity model is described in Ref. \cite{DHT}.

\subsubsection{Single relaxation time gradient model}

The BGK collision integral (\ref{BGKcoll}) has the following
important property: the linearization of the operator
(\ref{BGKcoll}) around the local equilibrium point has a very
simple spectrum $\{0,-1/\tau\}$, where $0$ is the ${n_{\rm
c}}$-times degenerate eigenvalue corresponding to the conservation
laws, while the non-zero eigenvalue corresponds to the rest of the
(kinetic) eigenvectors. Nonlinear collision operators which have
this property of their linearizations at equilibrium are called
single relaxation time models (SRTM). They play an important role
in modelling because they allow for the simplest identification of
transport coefficients.

The SRTM, based on the given entropy function $H$, is constructed
as follows (single relaxation time gradient model, SRTGM). For the
system with ${n_{\rm c}}$ local conservation laws, let $\ee_s$,
$s=1,\dots,{n_{\rm d}}-{n_{\rm c}}$, be an orthonormal basis in
the kinetic subspace, $\langle\mm_i|\ee_s\rangle=0$, and
$\langle\ee_s|\ee_p\rangle=\delta_{sp}$. Then the single
relaxation time gradient model is
\begin{equation}
\label{family} \bDelta=-\frac{1}{\tau}\sum_{s,p=1}^{{n_{\rm
d}}-{n_{\rm c}}}\ee_{s} K_{sp}(\ff) \langle\ee_{p}|\bnabla
H\rangle,
\end{equation}
where $K_{sp}$ are elements of a positive definite $({n_{\rm
d}}-{n_{\rm c}})\times({n_{\rm d}}-{n_{\rm c}})$ matrix $\KK$,
\begin{eqnarray}
\label{SRTM} \KK(\ff)&=&\CC^{-1}(\ff),\\\nonumber
C_{sp}(\ff)&=&\langle\ee_s|\bnabla\bnabla H(\ff) |\ee_{p}\rangle.
\end{eqnarray}
Here, $\bnabla\bnabla H(\ff)$ is the ${n_{\rm d}}\times {n_{\rm
d}}$ matrix of second derivatives, $\partial^{2}H/\partial
f_i\partial f_j$. Linearization of the collision integral at
equilibrium has the form,
\begin{equation}
\label{P} \LL=-\frac{1}{\tau}\sum_{s=1}^{{n_{\rm d}}-{n_{\rm
c}}}\ee_s\ee_s,
\end{equation}
which is obviously single relaxation time. Use of the SRTGM instead
of the BGK model results in the same hydrodynamics even when
the local equilibrium is not known in a closed form. Further details
of this model can be found in Ref. \cite{AK3}.

\subsubsection{$H$-functions of minimal kinetic models} \label{Hfunc}

The Boltzmann entropy function written in terms of the
one-particle distribution function $f({\xx}, \vel)$ is $H=\int
f\ln f \,d \vel$, where $\vel$ is the continuous velocity. Close
to the global (reference) equilibrium, this integral can be
approximated by using the Gauss--Hermite quadrature with the
weight
\[ W= (2\, \pi \, T_0)^{(D/2)}
\exp(-c^2/(2\,T_0)).\] Here $D$ is the spatial dimension, $T_0$ is
the reference temperature, while the particles mass and
Boltzmann's constant $k_{\rm B}$ are set equal to one. This gives
the entropy functions of the discrete-velocity models
\cite{DHT,AK4,AK5},
\begin{equation}
\label{app:H} H=\sum_{i=1}^{{n_{\rm d}}}
f_{i}\ln\left(\frac{f_{i}}{w_i} \right).
\end{equation}
Here, $w_i$ is the weight associated with the $i$-th discrete
velocity $\vel_i$ (zeroes of the Hermite polynomials).  The
discrete-velocity distribution functions (populations)
$f_i({\xx})$ are related to the values of the continuous
distribution function at the nodes of the quadrature by the
formula,
\[f_i({\xx})=w_i(2\, \pi \, T_0)^{(D/2)}
\exp(c^2_i/(2\,T_0))f({\xx}, {\vel}_i).\]  The  entropy functions
(\ref{app:H}) for various $\{w_i,\vel_i\}$ are the only input
needed for the construction of minimal kinetic models.

With the increase of the order of the Hermite polynomials used in
evaluation of the quadrature (\ref{app:H}), a better approximation
to the hydrodynamics is obtained. The first few models of this
sequence are represented in Table \ref{Tab: DiscV}.

\begin{table}[t]
  \caption{\label{Tab: DiscV} Minimal kinetic models.
Column 1: Order of Hermite velocity polynomial used to evaluate
  the Gauss-Hermite quadrature;
Column 2: Locally conserved (hydrodynamic) fields; Column 3:
Discrete velocities for $D=1$ (zeroes of the corresponding Hermite
polynomials). For $D>1$, discrete velocities are all possible
tensor products of the one-dimensional velocities in each
component direction; Column 4: Weights in the entropy formula
(\ref{app:H}), corresponding to the discrete velocities of the
Column 3. For $D>1$, the weights of the discrete velocities are
products of corresponding one-dimensional weights; Column 5:
Macroscopic equations for the fields of Column 2 recovered in the
hydrodynamic limit of the model.  }
   \bigskip
\begin{tabular}{|l|l|l|l|l|}
  \hline
  1. Order & 2. Fields    & 3. Velocities & 4. Weights & 5. Hydrodynamic limit \\
  \hline
  $2$ & $\rho$ & $ \sqrt{T_0}$ & $1/2$ & Diffusion   \\
      &        & $-\sqrt{T_0}$ & $1/2$ &             \\
  \hline
  $3$ & $\rho$, $\rho\uu$ & $0$                   & $2/3$ & Isothermal Navier--Stokes \\
      &                   & $\sqrt{3}\sqrt{T_0}$  & $1/6$ &
\\
      &                   & $-\sqrt{3}\sqrt{T_0}$ & $1/6$ &
  \\
  \hline
  $4$ &  $\rho$, $\rho\uu$, $e$ & $ \sqrt{3-\sqrt{6}}\sqrt{T_0}$ & $1/[4(3-\sqrt{6})]$ & Thermal Navier-Stokes \\
      &                         & $-\sqrt{3-\sqrt{6}}\sqrt{T_0}$ & $1/[4(3-\sqrt{6})]$ &                       \\
      &                         & $ \sqrt{3+\sqrt{6}}\sqrt{T_0}$ & $1/[4(3+\sqrt{6})]$ &                       \\
      &                         & $-\sqrt{3+\sqrt{6}}\sqrt{T_0}$ & $1/[4(3+\sqrt{6})]$ &                       \\
      \hline
\end{tabular}
\end{table}

\subsection{Entropic lattice Boltzmann method}

If the set of discrete velocities forms the links of a Bravais
lattice (with possibly several sub-lattices), then the
discretization of the discrete velocity kinetic equations in time
and space is particularly simple, and leads to the entropic
lattice Boltzmann scheme. This happens in the important case of
the isothermal hydrodynamics. The equation of the entropic lattice
Boltzmann scheme reads
\begin{equation}
\label{LBMdiscrete} f_i(\xx+\cc_i\delta t,t+\delta t)-f_i(\xx,t)=
\beta\alpha(\ff(\xx,t))\Delta_i(\ff(\xx,t)),
\end{equation}
where $\delta t$ is the discretization time step, and
$\beta\in[0,1]$ is a fixed parameter which matches the viscosity
coefficient in the long-time large-scale dynamics of the kinetic
scheme (\ref{LBMdiscrete}). The function $\alpha$ of the
population vector defines the maximal over-relaxation of the
scheme, and is found from the entropy condition,
\begin{equation}
\label{HTdiscrete}
H(\ff(\xx,t)+\alpha\bDelta(\ff(\xx,t))=H(\ff(\xx,t)).
\end{equation}
The nontrivial root of this equation is found for populations at
each lattice site. Equation (\ref{HTdiscrete}) ensures the
discrete-time $H$-theorem, and is required in order to stabilize
the scheme if the relaxation parameter $\beta$ is close to one. We
note in passing that the latter limit is of particular importance
in the applications of the entropic lattice Boltzmann method to
hydrodynamics because it corresponds to vanishing viscosity, and
hence to numerically stable simulations of  very high Reynolds
number flows.

\subsection{Entropic lattice BGK method (ELBGK)}

An important simplification occurs in the case of the isothermal
simulations when the entropy function is constructed using
third-order Hermite polynomials (see Table \ref{Tab: DiscV}): the
local equilibrium population vector can be obtained in closed form
\cite{AK5}. This enables the simplest entropic scheme -- the
entropic lattice BGK model -- for simulations of isothermal
hydrodynamics. We present this model in dimensionless lattice
units.

Let $D$ be the spatial dimension. For $D=1$, the three discrete
velocities are
\begin{equation}\label{1Dvel}
\cc=\{-1, 0, 1\}.
\end{equation}
For $D>1$, the discrete velocities are tensor products of the
discrete velocities of these one-dimensional velocities. Thus,
we have the $9$-velocity model for $D=2$ and the $27$-velocity
model for $D=3$. The $H$ function is Boltzmann-like:
\begin{equation}
\label{app:H27} H=\sum_{i=1}^{3^D} f_{i}\ln\left(\frac{f_{i}}{w_i}
\right).
\end{equation}
The weights $w_i$ are associated with the corresponding discrete
velocity $\vel_i$. For $D=1$, the three-dimensional vector of the
weights corresponding to the velocities (\ref{1Dvel}) is
\begin{equation}\label{weights1D} \ww = \left
\{\frac{1}{6},\frac{2}{3}, \frac{1}{6} \right \}.
\end{equation}
For $D>1$, the weights are constructed by multiplying the weights
associated with each component direction.

The local equilibrium minimizes the $H$-function (\ref{app:H})
subject to the fixed density and momentum,
\begin{equation}\label{Hyd27}
\sum_{i=1}^{3^D}f_i=\rho,\ \sum_{i=1}^{3^D}f_ic_{i\alpha}=\rho
u_{\alpha},\ \alpha=1,\dots,D.
\end{equation}
The explicit solution to this minimization problem reads,
\begin{equation}
\label{TED}
 f^{\rm eq}_i=\rho w_i\prod_{\alpha=1}^{D}
\left(2 -\sqrt{1+ 3 {u_{\alpha}^2}} \right)
\left(\frac{2\,u_{\alpha} + \sqrt{1+ 3\,u_{\alpha}^2}}{1-u_\alpha}
\right)^{c_{i\alpha}}.
\end{equation}
Note that the exponent, $c_{i\alpha}$, in (\ref{TED}) takes
the values $\pm 1, \, \mbox{and}\,\,  0$ only,
and the speed of sound, $c_{\rm s}$, in this model is equal to
$1/\sqrt{\,3}$. The factorization of the local equilibrium
(\ref{TED}) over spatial components is quite remarkable, and
resembles the familiar property of the local Maxwellians.

The entropic lattice BGK model for the local equilibrium
(\ref{TED}) reads,
\begin{equation}
\label{LBGKdiscrete} f_i(\xx+\cc_i\delta t,t+\delta t)-f_i(\xx,t)=
-\beta\alpha(f_i(\xx,t)-f_i^{\rm
eq}(\rho(\ff(\xx,t)),\uu(\ff(\xx,t))).
\end{equation}
The parameter $\beta$ is related to the  relaxation time $\tau$ of
the BGK model (\ref{BGKcoll}) by the formula,
\begin{equation}\label{relationtau}
\beta=\frac{\delta t}{2\tau+\delta t},
\end{equation}
and the value of the over-relaxation parameter $\alpha$ is computed
at each lattice site from the entropy estimate,
\begin{equation}
H(\ff-\alpha(\ff-\ff^{\rm eq}(\ff)))=H(\ff).
\end{equation}
In the hydrodynamic limit, the model (\ref{LBGKdiscrete})
reconstructs the Navier-Stokes equations with the viscosity
\begin{equation}
\label{viscELBGK} \mu=\rho c_{\rm s}^2\tau=\rho c_{\rm s}^2\delta
t\left(\frac{1}{2\beta}-\frac{1}{2}\right).
\end{equation}
The zero-viscosity limit corresponds to $\beta\to 1$.

\section{Wall boundary conditions} \label{BCsection}

The boundary (a solid wall) $\partial R$ is specified at any point
$ {\xx} \in\partial R $ by the inward unit normal $\bm{e}$, the
wall temperature $T_{\rm w }$, and the wall velocity ${\uu}_{\rm w
}$. The simplest boundary condition for the minimal kinetic models
is obtained upon evaluation of the diffusive wall boundary
condition for the Boltzmann equation \cite{Cerci} with the help of
the Gauss-Hermite quadrature \cite{AK4,AS,ELBMMICRO}. The essence
of the diffusive boundary condition is that particles loose their
memory of the incoming direction after reaching the wall. Once a
particle reaches the wall, it gets redistributed in a way
consistent with the mass-balance and normal-flux conditions.
Further, the boundary condition must also satisfy the condition of
detailed balance: if the incoming populations are at equilibrium
(corresponding to the wall-velocity), the outgoing populations are
also at equilibrium (corresponding to the wall-velocity).

\begin{figure}[ht]
 \begin{center}
 \includegraphics[scale=0.5]{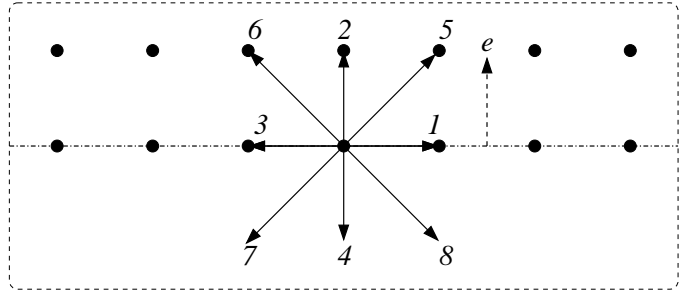}
 \end{center}
 \caption{\label{Fig1}
  Schematic diagram for the situation near a flat wall, when the
  wall normal, $\bm{e}$, (pointing towards the fluid) is in the
  positive $y$ direction.}
\end{figure}

For the purpose of simulations below, let us  consider the case
when the wall normal, $\bm{e}$, (pointing towards the fluid) is in
the positive $y$ direction. The lattice for the for the 9-velocity
isothermal model is depicted in Fig. \ref{Fig1}.

For this particular case, the boundary update rules for incoming
and grazing populations on a two-dimensional lattice are:
\begin{eqnarray}
\begin{aligned}
f_0(x,y, t+\delta t) &=f_0^{*}(x,y, t),\\
f_1(x,y, t+\delta t) &=f_0^{*}(x-c \delta t,y, t),\\
f_3(x,y, t+\delta t) &=f_0^{*}(x+c \delta t,y, t),\\
f_4(x,y, t+\delta t) &= \frac{1}{2}\left[f_4^{*}(x,y+c \delta t, t)+f_4^{*}(x, y,
  t)\right],\\
f_7(x,y, t+\delta t) &= \frac{1}{2}\left[f_7^{*}(x+c \delta t,y+c \delta t, t)+f_7^{*}(x, y,
  t)\right],\\
f_8(x,y, t+\delta t) &= \frac{1}{2}\left[f_8^{*}(x-c \delta t,y+c \delta t, t)+f_8^{*}(x, y,
  t)\right],
\end{aligned}
\end{eqnarray}
where $f^{*}$ denotes post-collision populations, and the update
rules for outgoing populations are:
\begin{eqnarray}
\begin{aligned}
f_2(\bx, t+ \delta t) &= f_2^{\rm eq}( \rho,\bu_{\rm wall})
\frac{f_4(\bx,t+ \delta t)+f_7(\bx,t+ \delta t)+f_8(\bx,t+ \delta t)}{f_2^{\rm eq}( \rho,\bu_{\rm
    wall})+f_5^{\rm eq}( \rho,\bu_{\rm wall})+f_6^{\rm eq}(
  \rho,\bu_{\rm wall})},\\
f_5(\bx, t+ \delta t) &= f_5^{\rm eq}( \rho,\bu_{\rm wall})
\frac{f_4(\bx,t+ \delta t)+f_7(\bx,t+ \delta t)+f_8(\bx,t+ \delta t)}{f_2^{\rm eq}( \rho,\bu_{\rm
    wall})+f_5^{\rm eq}( \rho,\bu_{\rm wall})+f_6^{\rm eq}(
  \rho,\bu_{\rm wall})},\\
f_6(\bx, t+ \delta t) &= f_6^{\rm eq}( \rho,\bu_{\rm wall})
\frac{f_4(\bx,t+ \delta t)+f_7(\bx,t+ \delta t)+f_8(\bx,t+ \delta t)}{f_2^{\rm eq}( \rho,\bu_{\rm
    wall})+f_5^{\rm eq}( \rho,\bu_{\rm wall})+f_6^{\rm eq}(
  \rho,\bu_{\rm wall})}.
\end{aligned}
\end{eqnarray}

\section{Simulation of quasi-two-dimensional flows\\ with
         two-dimensional kinetic models}
         \label{mapsection}

As mentioned in the introduction, many microflows of engineering
interest can be considered as quasi-two-dimensional. This means
that averaged quantities such as flow velocity and density do not
depend appreciably on the third spatial direction. Thus, it is
tempting to use two-dimensional kinetic models in simulations of
such flows. However, care should be taken in order to map
correctly the results of two-dimensional simulations onto
experimental data or molecular dynamics simulations. Indeed,
molecular motion remains three-dimensional in spite of the fact
that some averages can be considered  two-dimensional. In the DSMC
simulations of two-dimensional flows, for example, collisions of
the particles are always treated as three-dimensional. The
two-dimensional kinetic models therefore must be considered as a
computational device which uses fictitious particles moving in two
dimensions in order to mimic quasi-two-dimensional flows of
particles moving in three-dimensions.

The mapping of the parameters of the three-dimensional kinetic
equation onto the two-dimensional lattice Boltzmann scheme is done
in two steps:
\begin{itemize}
\item Map the continuous $3D$ kinetic equation onto the $3D$
discrete velocity model. \item Map the $3D$ discrete velocity
model onto the $2D$ velocity model.
\end{itemize}

In the case considered in this paper, the $3D$ continuous kinetic
model is the BGK model \cite{Cerci}, which contains the relaxation
parameter $\tau_{\rm BGK}$,
\begin{equation}
\label{contBGK}
    \partial_t f+c_{\alpha}\partial_{\alpha}f=-\frac{1}{\tau_{\rm
    BGK}}(f-f^{\rm LM}),
\end{equation}
where $f^{\rm LM}$ is the local Maxwell distribution function. In
this subsection we shall explicitly indicate all the functions and
parameters related to the continuous BGK model with the subscript
in order to distinguish them from the lattice counterparts.

The viscosity coefficient of the BGK model (\ref{contBGK}) is
related to the relaxation time $\tau_{\rm BGK}$ as follows:
%
\begin{equation}
\label{viscBGK} \mu_{\rm BGK}=\rho_{\rm BGK} T_{\rm BGK} \tau_{\rm
BGK}.
\end{equation}

In the isothermal discrete velocity BGK model of section
\ref{ELBMsec}, the viscosity is expressed not as a function of
temperature but of sound speed $c_{\rm s}$,
\begin{equation}
\mu=\rho c_{\rm s}^2\tau,
\end{equation}
where $c_{\rm s}=1/\sqrt{3}$ in lattice units used in simulations.
Therefore, in the first step of the above procedure, we map the
parameters of the  continuous $3D$ BGK equation onto the $3D$
discrete velocity model using the relation for the speed of sound,
$c_{\rm s\ BGK}=\sqrt{\gamma T_{\rm BGK}}$, where $\gamma$ is the
adiabatic exponent. For an ideal gas, $\gamma=5/3$. Thus, the
first step  is accomplished by the relation,
\begin{eqnarray}
 c_{\rm s\ BGK}&=&c_{\rm s},\nonumber\\
 \sqrt{\frac{5}{3} T_{\rm BGK}}&=&\sqrt{\frac{1}{3}}.\label{map1}
\end{eqnarray}
This formula establishes the relation between the relaxation
parameter of the  $3D$ continuous BGK equation (\ref{contBGK}) and
the $3D$ isothermal discrete velocity model of section
(\ref{ELBMsec}). Note that the sound speed of the thermal model is
made equal to the sound speed of the isothermal model by the
relation (\ref{map1}).

At the second step, we map the three-dimensional 27-velocity
isothermal model onto the two-dimensional 9-velocity model. Note
that the sound speed in both models is {\it identical} (and equals
to $\sqrt{k_{\rm B}T_0}$ in dimensional units). The mapping is
done by populating at the equilibrium the links of the 27-velocity
lattice in the direction orthogonal to the fixed plane containing
the links of the 9-velocity sub-lattice. This amounts to the
following recomputation of the three-dimensional density
$\rho_{3D}$ in terms of the two-dimensional density $\rho_{2D}$:
\begin{equation}
\rho_{3D}=\frac{3}{2}\rho_{2D}.
\end{equation}

This formula enables the computation of the effective three-dimensional
density in terms of the two-dimensional density used in the
simulations with the 9-velocity model. Specification of density
is an important part of the simulation since it is used to define
data as prescribed pressure drop at the inlet and outlet of pipes etc.

The formulas collected in this section, together with the
viscosity formula of the fully discretized entropic lattice
Boltzmann method (\ref{viscELBGK}), enable the comparison of
two-dimensional simulation results with the results obtained by
microscopic simulations with different collision models (the hard
sphere model, for example). The choice of the model is needed to
identify the mean free path which is is model-dependent,
especially at moderate values of Knudsen number.

\section{Plane Poiseuille flow} \label{Simsec}

Plane Poiseuille flow is one of the most studied benchmarks on gas
dynamics. The gas moves between two parallel plates driven by a
fixed pressure difference between the inlet and outlet. It is well
known that for this setup the flow rate through a cross-section of
the pipe exhibits a minimum \cite{Knudsen, Cerci}. In fact, one of
the major achievements in the early days of kinetic theory was the
prediction of a minimum of the mass flow rate as a function of the
Knudsen number for $Kn \sim 1$.

We simulate the two-dimensional flow in a rectangular duct of
length $L$ along the streamwise direction ($x$) and width $H\ll L$
along the wall-normal ($y$) direction. The flow is driven by a
fixed pressure difference $\Delta P = P_{\rm in}- P_{\rm out}$,
where $P_{\rm in}$ and $P_{\rm out}$ are the pressure at the inlet
and outlet of the duct, respectively. In the subsequent analysis,
we shall follow the convention used by Cercignani, where the
Knudsen number for the BGK model is defined as (continuous BGK
units):
\begin{equation}
  {\rm Kn} = \frac{ \mu \sqrt{2 T_0}}{P_0H},
\end{equation}
where the pressure $P_0$ is defined as the mean of the inlet and
outlet pressures, $P_0=(P_{\rm in}+P_{\rm out})/2$. In the
hydrodynamic limit, this results in the well known
Hagen-Poiseuille parabolic velocity profile:
\begin{equation}
u(y) = U_0 \left( \frac{1}{4} -  \frac{y}{H}^2 \right),
\end{equation}
where the amplitude of the flow for a two-dimensional duct is
$U_0 = H^2 \Delta P/ (2 \mu \,L )$. From the  analysis of the
Boltzmann-BGK equation it is known that the dimensionless flow rate
\begin{equation}
Q = \frac{1}{H U_0 {\rm Kn }} \int _{H/2}^{H/2}  u(y) dy,
\end{equation}
has a low-Knudsen $({{\rm  Kn } \ll 1})$ asymptotic:
\begin{equation}
Q_0 =\frac{1}{6 {\rm Kn}}+s+(2s^2-1)\;{\rm Kn}  ,
\end{equation}
with $s=1.015$ (see \cite{Cerci}), which is equivalent to solving
the Navier-Stokes equation with a second-order slip boundary
condition. Further, the high-Knudsen asymptotic expression for the
flow rate is:
\begin{eqnarray}
  Q_{\infty} &\sim& \frac{1}{\sqrt{\pi}} \; \log{({\rm Kn})} + O(1),
\end{eqnarray}
which implies a slow logarithmic divergence of the flow rate as a
function of the Knudsen number. These two asymptotic limits ensure
that the flow rate must have at least one minimum  at some finite
$\rm Kn$. The low-Knudsen asymptotic behavior is related to the
situation where on the average the number of collisions
encountered by a molecule in the bulk is much larger than the
collisions with the wall. Thus, the effective balance between the
frictional forces (due to collisions) and the applied pressure
gradient ensures a parabolic velocity  profile (with a slip at the
wall).  On the other hand, any molecular motion in high Knudsen
number flows is retarded mostly by collisions occurring at the
wall. This leads to an effectively flat velocity profile in the
channel.

In addition to the asymptotic analysis, more detailed
investigations of the linearized Boltzmann equation are available
(see, for example, \cite{CerciVar}). In the next subsection, we
compare the result of our numerical simulation with the numerical
result of \cite{CerciVar} and the asymptotic expressions.

\subsection{Numerical simulation}

\begin{figure}[ht]
 \begin{center}
 \includegraphics[scale=0.6]{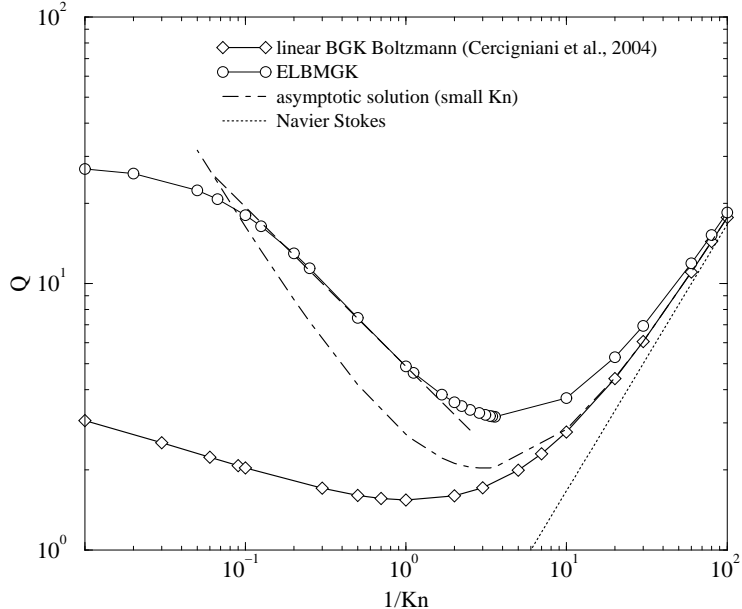}
 \end{center}
 \caption{\label{Fig2}
Comparison of the ELBM solution for the dimensionless flow rate
with the low $\rm Kn$ asymptotic solution, and the numerical
solution of the stationary linearized BGK equation
\cite{CerciVar}. The dashed line indicates the qualitative
agreement with the expected logarithmic scaling at higher $\rm
Kn$. All the curves become indistinguishable at ${\rm Kn}\leq
0.01$.}
 \end{figure}

Poiseuille flow in a two-dimensional duct was simulated with the
entropic LBGK model for a range of Knudsen numbers. The length to
height ratio used in the simulation was $30$, and the resolution
was taken $1101 \times 45$ points. In lattice units, the Knudsen
number is:
 \begin{equation}
{\rm Kn} =  \frac{\tau}{H}\sqrt{\frac{2}{5}}.
\end{equation}
The simulation results for the flow-rate is plotted in Fig.
\ref{Fig2} together with the numerical solution of the stationary
linearized BGK model \cite{CerciVar} and the asymptotic
expressions. Some of the important conclusions following from this
study are:
\begin{itemize}
\item Quantitative agreement with the fully microscopic results is
obtained in the range of $0 \le Kn \le 0.01$. This indicates that
the {\it parameter-free} ELBGK model can be used in the domain of
slip-flow for quantitative simulations.

\item At higher Knudsen numbers, the simulation results predict
the expected logarithmic divergence (as indicated by the dashed
line in Fig.\ ref{Fig:Flowrate}), but they do deviate from the
numerical solution of the linearized stationary BGK equation. At
very high Knudsen numbers, the finite size of the duct will always
lead to a nearly flat (independent of $\rm Kn$) flow rate.

\item Our simulations predict the Knudsen minimum at moderate
values of $\rm Kn$. The minimum is found around $\rm Kn=0.35$,
whereas much more elaborate (and computationally expensive)
molecular dynamics simulations \cite{Karniadakis2} and the
semi-analytical solutions to the stationary BGK equation indicate
the location of the minimum at $\rm Kn$ in the range $0.8-1$. The
discrepancy is not surprising because the model used in our
simulation is drastically simpler than any of the alternative
models. In particular, the built-in isothermal assumption limits
its domain of validity to the cases when incompressibility is a
good approximation indeed (slip-flow). In the domain of $\rm
Kn\sim 1$ it is therefore required to use the non-isothermal model
which includes correctly the energy redistribution in the
collision processes.

\end{itemize}

\section{Discussion and conclusions} \label{conclusion}

In this paper, we have set up the basis of a new computational
approach for the simulation of microflows - the entropic lattice
Boltzmann method. We have described the hierarchy of entropic
lattice Boltzmann models for both isothermal and thermal
simulations. Here, we have focused on the simplest isothermal
model - the entropic lattice BGK equation and presented in detail
the implementation of the diffusive wall boundary condition in the
lattice Boltzmann setting. For the ELBGK model, we derived the
parameter map of two-dimensional isothermal simulations on the
three-dimensional data, and have tested all this on the benchmark
problem of the planar Poiseuille flow in the full range of
rarefaction (Knudsen number) from the nearly-continuous case to
the free-molecular flow. Thus, we confirm the validity of the
entropic lattice Boltzmann method as a viable tool for
computations of the microflows in the most relevant to MEMS
applications domain of $\rm Kn$ up to ${\rm Kn}\sim 0.1$. This
confirmation clearly points to the usefulness of development of
more sophisticated lattice Boltzmann models in this range of
parameters which is the subject of further studies. It should be
stressed that applications of the isothermal models can be
considered as a parameter-free slip-flow hydrodynamic models, and
that the way to extend the domain of the quantitative predictions
must take into account energy conservation in the collisions. We
would also like to point out the computational efficiency of the
proposed models. For example, the full data set presented in Fig.\
\ref{Fig2} was computed within several hours on a single-processor
computer facility.

Finally, let us briefly mention the following by-product of our
study, relevant for the use of the lattice Boltzmann simulations
in molecular models. In particular, the Direct Simulation Monte
Carlo method (DSMC) requires a good initial choice of the velocity
distribution function to perform efficiently. Most of the current
simulations use the local equilibrium (or equilibrium) Maxwell
distribution function. A better choice could be the anisotropic
Gaussian approximation
\begin{equation}
\label{anisotropic} f(\vv,\xx)\sim
\exp\left(-(v_{\alpha}-u_{\alpha})\rho
P^{-1}_{\alpha\beta}(v_{\beta}-u_{\beta})\right),
\end{equation}
where $P^{-1}$ is the inverse of the pressure tensor,
\begin{equation}
P_{\alpha\beta}=\int
\frac{(v_{\alpha}-u_{\alpha})(v_{\beta}-u_{\beta})}{2}f d\vv.
\end{equation}

The LB data for the stationary stress tensor and for the flow
field can be used as initial guess for the functions
$u_{\alpha}(\xx)$ and $P^{-1}_{\alpha\beta}(\xx)$ in the velocity
distribution function (\ref{anisotropic}). This initial condition
for the DSMC simulation of the stationary will dramatically reduce
the simulation time as compared to the case when it is initialized
at equilibrium.

{\it Acknowledgments.} This work was supported by the Bundesamt
f\"ur Energie,  BFE-Project Nr. 100862 "Lattice Boltzmann
simulations for chemically reactive systems in a micrometer
domain". Discussions with A.\ N.\ Gorban, S.\ Succi, A.\
Tomboulides and F. Toschi are kindly acknowledged.

\bibliographystyle{unsrt}
\bibliography{KnudMin}
\end{document}